\documentclass[12pt,preprint]{aastex}

\slugcomment{Not to appear in Nonlearned J., 45.}

\shorttitle{X-rays from Gl\,569\,Bab}
\shortauthors{Stelzer}

\begin{document}


\title{Quiescent X-ray Emission from an evolved Brown Dwarf ?}


\author{B. Stelzer\altaffilmark{1,2}}
\altaffiltext{1}{Dipartimento di Scienze Fisiche ed Astronomiche, Universit\`a di Palermo, Piazza del Parlamento 1, I-90134 Palermo}
\altaffiltext{2}{INAF - Osservatorio Astronomico di Palermo, Piazza del Parlamento 1, I-90134 Palermo}

\email{stelzer@astropa.unipa.it}




\begin{abstract}
I report on the X-ray detection of Gl\,569\,Bab. During a $25$\,ksec {\em Chandra} observation 
the binary brown dwarf is for the first time spatially separated in X-rays from the flare star 
primary Gl\,569\,A. 
Companionship to Gl\,569\,A constrains the age of the brown dwarf pair to $\sim 300-800$\,Myr.  
The observation presented here is only the second X-ray detection of an evolved brown
dwarf. About half of the observing time is dominated by a large flare on 
Gl\,569\,Bab, the remainder is characterized by weak and non-variable emission just above the
detection limit. This emission -- if not related to the afterglow of the flare -- 
represents the first detection of a quiescent corona on a brown dwarf, representing an important
piece in the puzzle of dynamos in the sub-stellar regime.  
\end{abstract}





\section{Introduction}

Late-type stars have long been known to display signatures of magnetic activity evidencing 
solar-like dynamo action \citep{Rosner85.1}. 
The picture is less clear for the coolest dwarf stars at the bottom 
of the main-sequence (MS) and in the substellar regime. These objects are fully convective
and can not drive a solar-like $\alpha\Omega$-dynamo. The efficiency of 
alternative field generating mechanisms, e.g. an $\alpha^2$-dynamo \citep{Raedler90.1} 
or a turbulent dynamo \citep{Durney93.1}, 
are both difficult to predict and observationally poorly constrained. 
Systematic investigations of chromospheric H$\alpha$ activity in very low-mass (VLM) stars and 
brown dwarfs (BDs) in the solar neighborhood \citep{Gizis00.1,Mohanty03.1} 
have shown that H$\alpha$ emission reaches a maximum at 
spectral type M7, and fades off for the latest M dwarfs suggesting a change in
the driving dynamo mechanism. 
X-ray emission is a complementary activity indicator, probing 
the effects of the magnetic processes in the outermost and hottest part of the
atmosphere, the corona.
Beyond spectral type M6 the X-ray regime is widely unexplored.
Only few late M dwarfs have been observed with {\em Chandra} and/or {\em XMM-Newton}:  
APMPM2354 \citep{Scholz04.1},  
vB\,10 \citep{Fleming03.1}, LP944-20 \citep{Rutledge00.1}. 
With the exception of vB\,10 all X-ray detections of M dwarfs cooler than spectral type M7 
are ascribed to flares, i.e. outbursts on a time-scale of hours that arise
from reconnection of magnetic field lines. It is unclear to date whether ultra-cool
dwarfs can sustain persistent X-ray emission, as is typically observed on higher-mass stars. 
In particular, a `quiescent' corona could not be established for any BD so far. 

While the observational material on X-ray activity
of evolved ultra-cool dwarf stars and BDs is scarce,
X-ray emission has lately been revealed from a substantial number of $\sim 1-5$\,Myr-old 
VLM pre-MS stars and BDs in various star forming regions 
\citep{Neuhaeuser98.1, Preibisch02.1, Mokler02.1, Feigelson02.1, Stelzer04.1}, 
for one BD in the $\sim 10-30$\,Myr-old 
TW\,Hya association \citep{Tsuboi03.1}, one member of Upper Sco \citep{Bouy04.1},
and one BD in the Pleiades \citep{Briggs04.1}.
BDs are not able to fuse hydrogen and subsequently must cool down and become fainter 
as they age. Therefore, their detection at young ages but non-detection in more evolved stages
could be due to: 
(A) a drop of the X-ray luminosity as a result of the decreased bolometric luminosity.  
A decline of $L_{\rm x}$ with age is expected if the $L_{\rm x}/L_{\rm bol}$ correlation 
typical for the higher-mass late-type stars holds also in the sub-stellar regime; 
(B) an effect of the atmospheric temperature.
Based on the absence or low level of H$\alpha$ emission seen in most
field L dwarfs \citet{Mohanty02.1} have argued that the
dynamo may shut off below a critical
temperature because the atmosphere becomes too neutral to provide substantial
coupling between matter and magnetic field. But due to a lack of systematic
observations this hypothesis has not been tested so far in the X-ray regime,
nor has a firm relation between X-ray and H$\alpha$ emission been established.
                                                                                
For a study of the evolution of X-ray activity on sub-stellar objects as they 
evolve away from the youngest ages BD companions to MS stars are favorable targets.
The age determination in the coolest part of the HR diagram is subject 
to considerable uncertainty because of a degeneracy in the mass-age plot and a 
uncertainty of theoretical evolutionary models for VLM stars and BDs. 
But ages of MS stars can be estimated by a number of methods 
\citep{Lachaume99.1}, and companions are expected to be coeval.         
                                                                    
A low-mass companion to the nearby ($9.8$\,pc; Dahn et al. 2002) dM2e star Gl\,569
was identified and confirmed as a common proper motion pair by \citet{Forrest88.1}. 
\citet{Martin00.1} found that the companion itself is a binary
of $0.1^{\prime\prime}$ separation.
Various methods have been used to constrain the age of Gl\,569\,A:
H$\alpha$ and X-ray activity \citep{Henry90.1, Pallavicini90.1}, 
the non-detection of Lithium absorption \citep{Magazzu93.1},
and its space velocity suggesting membership to the UMa association or a new super-cluster 
\citep{Chereul99.1}.
All these methods place it at an age between $300-800$\,Myr,  
consistent with the position of the  Gl\,569\,Bab binary in the HR diagram
\citep{Lane01.1}.
The dynamical masses of Gl\,569\,Ba and Gl\,569\,Bb were derived by 
\citet{Zapatero04.1} from a combination of radial velocity measurements and 
astrometry, yielding 99\,\% confidence intervals of 
$M_{\rm Ba} = 0.055-0.087\,M_\sun$ and $M_{\rm Bb} = 0.034-0.070\,M_\sun$.
The lower-mass component is thus the first object confirmed as substellar 
independent of models. 
In this Letter I present the {\em Chandra} detection of Gl\,569\,Bab, including
a large flare and probably the first detection of quiescent X-ray emission from 
an evolved BD.

\section{Observations}

The Gl\,569 system was observed with the Advanced CCD Imaging Spectrometer (ACIS),
using the S3 chip in imaging mode for a total exposure time of $25$\,ks. 
The data analysis was carried out using the CIAO software 
package\footnote{CIAO is made available by the CXC and can be downloaded 
from \\ http://cxc.harvard.edu/ciao/download-ciao-reg.html} version 2.3
in combination with the calibration database (CALDB) version 2.21.
I started the analysis with the level\,1 events file provided by the
pipeline processing at the {\em Chandra} X-ray Center (CXC). 
In the process of converting the level\,1 events file to a level\,2 events file
for each of the observations the following steps were performed: 
Charge transfer inefficiency (CTI) correction, 
removal of the pixel randomization, 
filtering for event grades (retaining the standard {\em ASCA} grades $0$, $2$, $3$, $4$, and $6$), 
and application of the standard good time interval (GTI) file. 
The events file was also checked for any systematic aspect offsets using 
CIAO software, but none were present.

\section{Data Analysis and Results}

\begin{figure}
\epsscale{.90}
\plotone{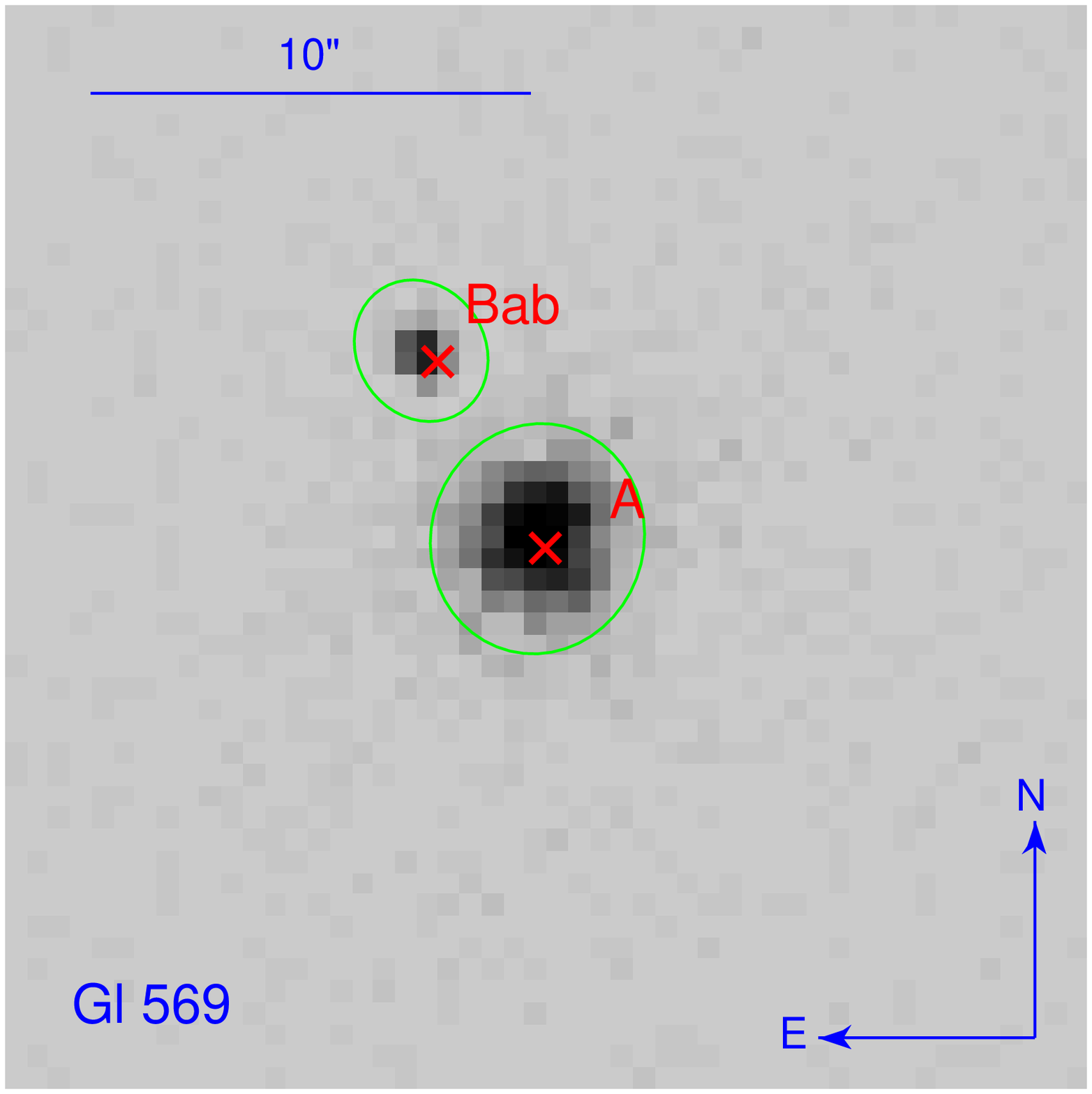}
\figcaption{{\em Chandra} ACIS-S image of the Gl\,569 system. The size of the image is $50 \times 50$~pixels. X-points denote the optical/IR position of Gl\,569\,A and Gl\,569\,Bab. Ellipses mark the corresponding X-ray sources.}
\label{fig:acis_image}
\end{figure}

The expected position of Gl\,569\,A was found by 
translating its {\em Hipparcos} position to the time of the {\em Chandra} observation
with help of the proper motion given by \citet{King03.1}. The position of Gl\,568\,Bab 
relative to the primary was obtained using the separation and P.A. of \citet{Lane01.1}. 
Source detection was carried out 
with the {\it wavdetect} algorithm 
on an image of $50 \times 50$\,pixels length ($1$\,pixel = $0.492^{\prime\prime}$)
centered on the computed position of the primary 
using wavelet scales between $1$ and $8$ in steps of $\sqrt{2}$. 
Two X-ray sources are found which are unambiguously identified with Gl\,569\,A 
(offset between optical and X-ray position $\Delta_{\rm ox} = 0.28^{\prime\prime}$) 
and Gl\,569\,Bab ($\Delta_{\rm ox} = 0.45^{\prime\prime}$), respectively; 
see Fig.~\ref{fig:acis_image}. 
Using the position of the X-ray source
corresponding to Gl\,569\,A as the reference, the second X-ray source is 
characterized by an angular separation of $(5.02 \pm 0.06)^{\prime\prime}$
and a position angle of $(31.7 \pm 0.6)^\circ$. 
These values are in reasonable agreement with those published by \citet{Lane01.1} 
based on near-IR images, the difference possibly being due to orbital motion of
the BD pair around the primary in the time interval of $\sim 3.4$\,yr.

I extracted the source photons of Gl\,569\,Bab from a circular area of $1.5^{\prime\prime}$ radius 
centered on the position of the corresponding X-ray source. 
According to a simulation of the point spread function (PSF) with {\em mkpsf} 
this area contains about $95$\,\% of the source photons assuming a photon 
energy of $1.5$\,keV. To eliminate contamination by counts
from the wings of the PSF of the X-ray bright primary the background was computed 
within an annulus centered on Gl\,569\,A and at the distance of the BD binary,
but excluding the source extraction area of Gl\,569\,Bab. 
After correcting for the missing $5$\,\% of photons the total number of source counts is $250$. 

The primary Gl569\,A suffers from pile-up, such that the flux is underestimated and the shape of the lightcurve
and the spectrum are distorted when source photons are extracted from a circular area 
centered on the X-ray source. 
At the expense of reduced statistics the negative effects of pile-up can be avoided if photons from 
the core of the PSF are ignored. 
To determine the optimum inner radius $r_{\rm in}$ for an annular source extraction region 
I extracted events from a series of annuli with $r_{\rm in} = 0.5^{\prime\prime}....2.25^{\prime\prime}$. 
The outer radius was fixed at $3.5^{\prime\prime}$ to exclude photons from the BD pair. 
With help of the ChaRT\footnote{Information on the usage of the Chandra Ray Tracer (ChaRT) can be found at http://cxc.harvard.edu/chart/} 
and MARX\footnote{The Model of AXAF Response to X-rays (MARX) is provided by the MIT/CXC (see http://space.mit.edu/CXC/MARX/)} simulators 
an individual {\em arf} was constructed for each of 
the annuli to correct the effective area for the missing part of the 
PSF\footnote{see http://www.astro.psu.edu/users/tsujimot/arfcorr.html for more information on
the procedure}. 
Then I examined how the observed luminosity, derived from the photon spectrum, 
changes depending on $r_{\rm in}$. 
The optimized extraction area was then defined as the annulus where $L_{\rm x}$ reaches a plateau,
$r_{\rm in}=1.25^{\prime\prime}$. 

\begin{figure}
\epsscale{1.0}
\plotone{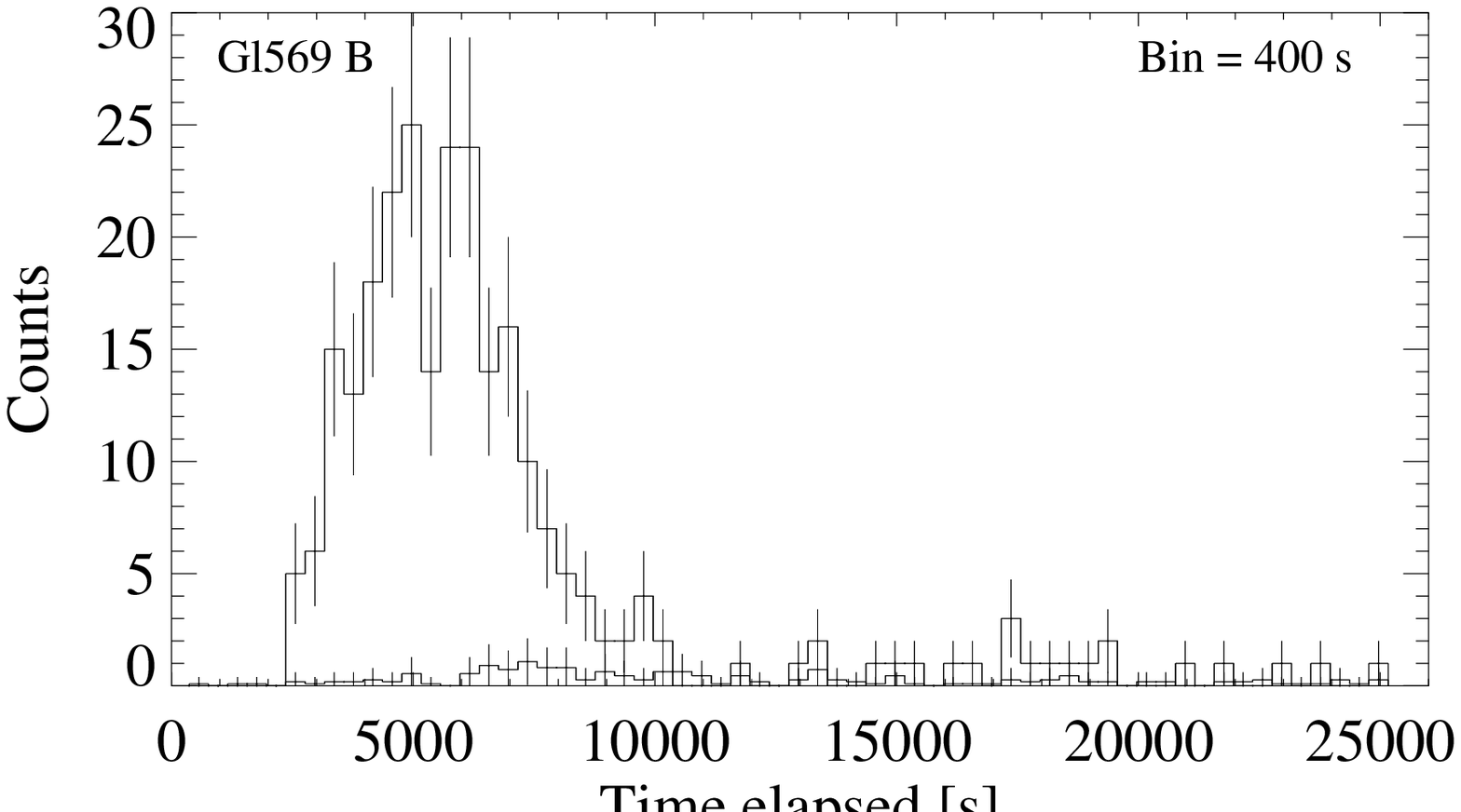}
\plotone{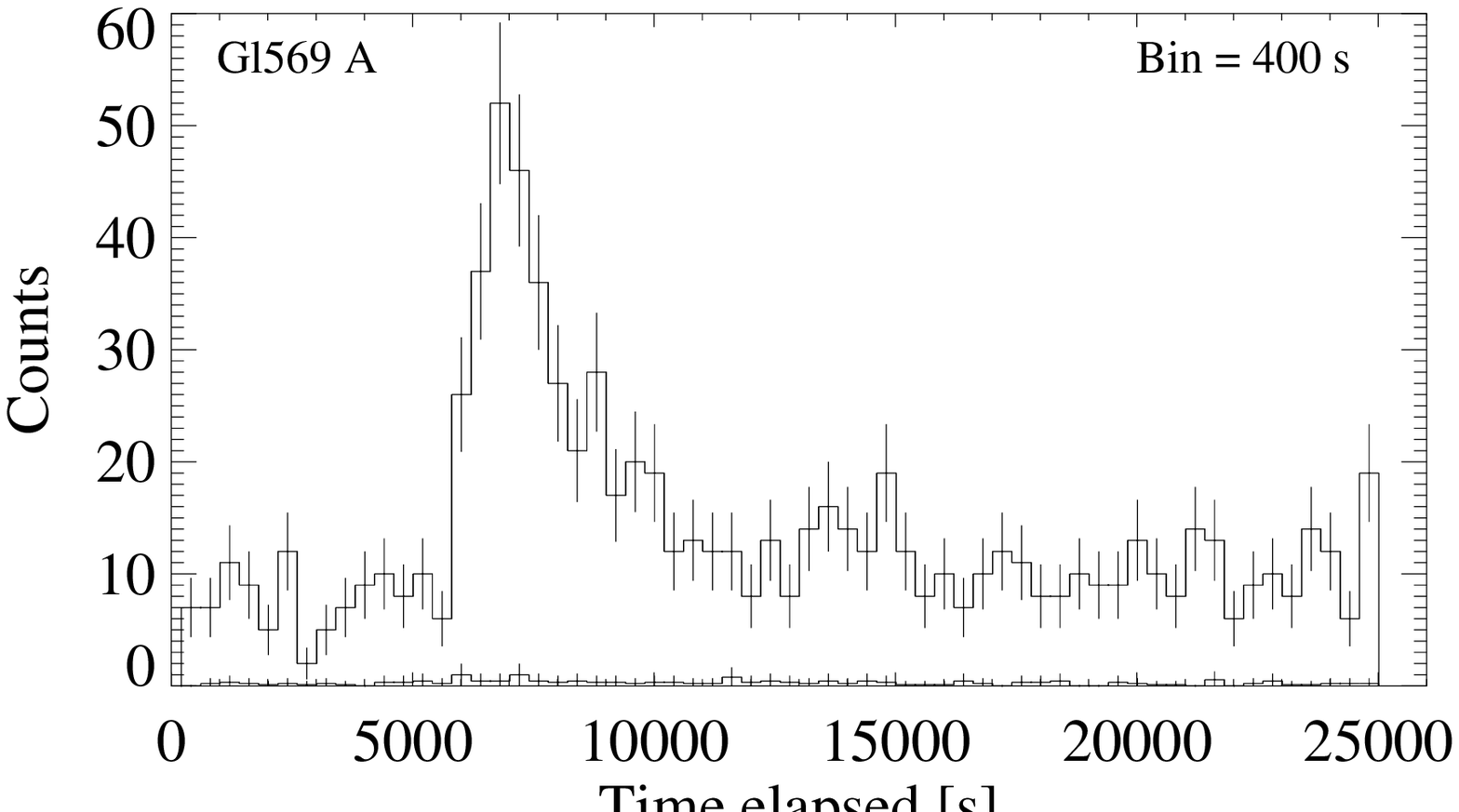}
\figcaption{Source and background lightcurves of Gl\,569\,Bab and Gl\,569\,A.}
\label{fig:acis_lc}
\end{figure}

The X-ray lightcurves of Gl\,569\,Bab and Gl\,569\,A are shown in Fig.~\ref{fig:acis_lc}. 
Both the BD pair and the M2 star underwent a spectacular flare.  
Note that the relative amplitude of the two lightcurves does not reflect the actual strength
of the outbursts because -- as explained above -- 
for Gl569\,A photons from only a small fraction of the PSF 
($4.5$\,\% for photon energy of $1.4967$\,keV) could be considered in the source 
extraction region. The strength of pileup depends on the changing incident photon flux. 
Therefore the exact time evolution of the intensity of Gl569\,A is difficult to reconstruct. 
The X-ray emission of Gl\,569\,A is not the main scope of this Letter, and will be discussed 
elsewhere. In the following I focus on Gl\,569\,Bab.

\section{X-ray Properties of Gl\,569\,Bab}

\begin{figure}
\epsscale{1.0}
\plotone{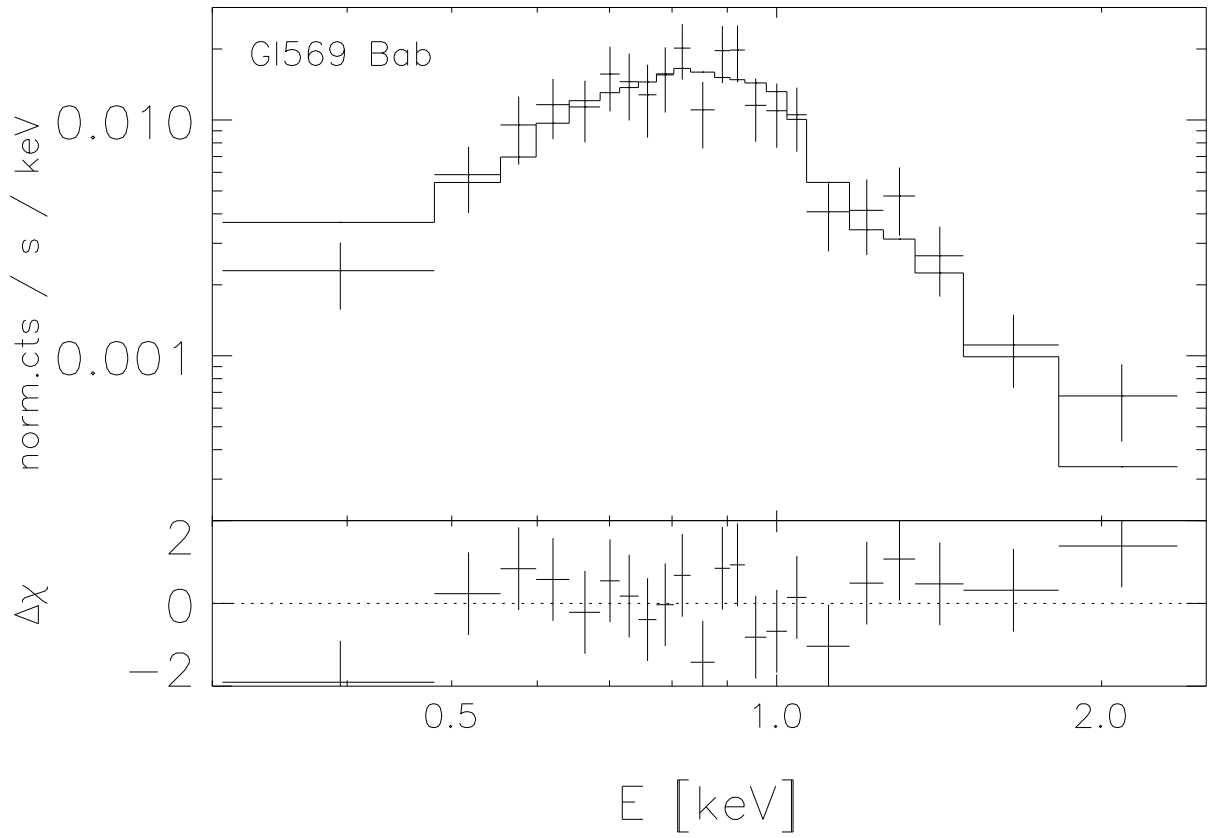}
\figcaption{X-ray spectrum of Gl\,569\,Bab, 2-T APEC model, and residuals.}
\label{fig:spec_gl569bab}
\end{figure}

The beginning of the observation is dominated by a strong flare on the BD binary, 
with $\sim 93$\,\% of all source photons being concentrated in the first $12$\,ksec of the
observation.  
In the remainder of the observing time a total of $23$ photons were collected in the source
area, of which about $6$ are expected to be background events, mostly contaminants
from Gl\,569\,A. This corresponds to a 
`quiescent' source count rate of $(1.2 \pm 0.3)\,10^{-3}$\,cps. Assuming a 1-T plasma 
without absorption yields $\log{L_{\rm Q}} = 25.8$\,erg/s. This value is 
insensitive to temperature within a typical range of $0.3...1.2$\,keV. 
The peak count rate during the flare is $\sim 0.0625$\,cps, amounting to a peak luminosity of
$\log{L_{\rm F}} = 27.5$\,erg/s using the same assumptions for the spectral shape as for the 
estimate of the `post-flare' luminosity. 
A more reliable estimate for the X-ray luminosity can be obtained from an analysis of the
X-ray spectrum. 
It provides information on the time-averaged luminosity, which is dominated by the flare. 
A statistically acceptable interpretation in terms of thermal emission requires a model 
with a minimum of two temperatures 
($kT_1=0.38^{+0.22}_{-0.15}$\,keV, $kT_2=0.94^{+0.28}_{-0.23}$\,keV, $EM_1/EM_2=1.2^{+1.6}_{-0.9}$).
Discrepancies between model and data suggest that absorption may not be negligible 
(see Fig.~\ref{fig:spec_gl569bab}). However,
if an absorption term is added, the fit results in very high emission measure for the soft component.
A correlation in the parameter space between $N_{\rm H}$ and $EM_1$ is typical for 
low-sensitivity, low-resolution spectra. 
The model without absorption seems to be more realistic, because (i) no visual extinction has been reported
for Gl\,569, and (ii) a dominant soft component is unexpected in flares.

\clearpage
\begin{table}
\begin{center}
\caption{X-ray luminosity and $L_{\rm x}/L_{\rm bol}$ ratio for Gl\,569\,Bab.}
\label{tab:lx_gl569bab}
\begin{tabular}{lcccc}\hline
                  &            & Quies         & Peak          & Average   \\ \hline 
$\log{L_{\rm x}}$ & [erg/s]    & $25.8$        & $27.5$        & $26.8$        \\
$\log{(L_{\rm x}/L_{\rm bol})_{\rm Ba}}$ & & $-4.4$ & $-2.7$ & $-3.4$ \\ 
$\log{(L_{\rm x}/L_{\rm bol})_{\rm Bb}}$ & & $-4.2$ & $-2.5$ & $-3.2$ \\ \hline  
\end{tabular}
\end{center}
\end{table}
\clearpage

A widespread used measure for coronal activity is the fraction
of X-ray luminosity relative to the bolometric luminosity. The two components in the BD binary have 
$\log{L_{\rm Ba}}\,{\rm [erg/s]} = 30.2$ and  $\log{L_{\rm Bb}}\,{\rm [erg/s]} = 30.0$ \citep{Gorlova03.1}.
Table~\ref{tab:lx_gl569bab} presents a summary of X-ray luminosities and 
$L_{\rm x}/L_{\rm bol}$ ratios for Gl\,569\,Bab derived from this {\em Chandra}
observation. $L_{\rm x}/L_{\rm bol}$ was evaluated attributing all observed X-rays to either of the 
unresolved components. Recall that, `Quies' and `Peak' values are estimated from count rates using model 
assumptions. `Average' denotes the luminosity derived from the analysis of the time averaged spectrum.

\section{Discussion}

I report on a huge X-ray flare and the possible detection of quiescent emission on Gl569\,Bab.  
The present $3\sigma$ limits on the individual masses do not exclude that the higher-mass component 
is a star at the borderline to the substellar regime, but most evidence points at a BD-BD binary 
(see discussion in Zapatero Osorio et al. 2004). 
Independent of whether the X-ray source is substellar or not, the emitter is one of only two 
ultra-cool dwarfs older than $\sim 100$\,Myr for which X-ray emission could be detected so far. 
Therefore, Gl\,569\,Bab constitutes an important link between activity in star forming
regions, associations, and open clusters, and the more evolved field stars and BDs. 
Gl\,569\,Bab are spectroscopically confirmed companions to a MS star. This eliminates 
two important parameters (the effective temperature and the age) from the list of unknowns, 
whose influence on the efficiency of (sub-)stellar dynamos has remained elusive so far.   

\begin{figure}
\epsscale{1.0}
\plotone{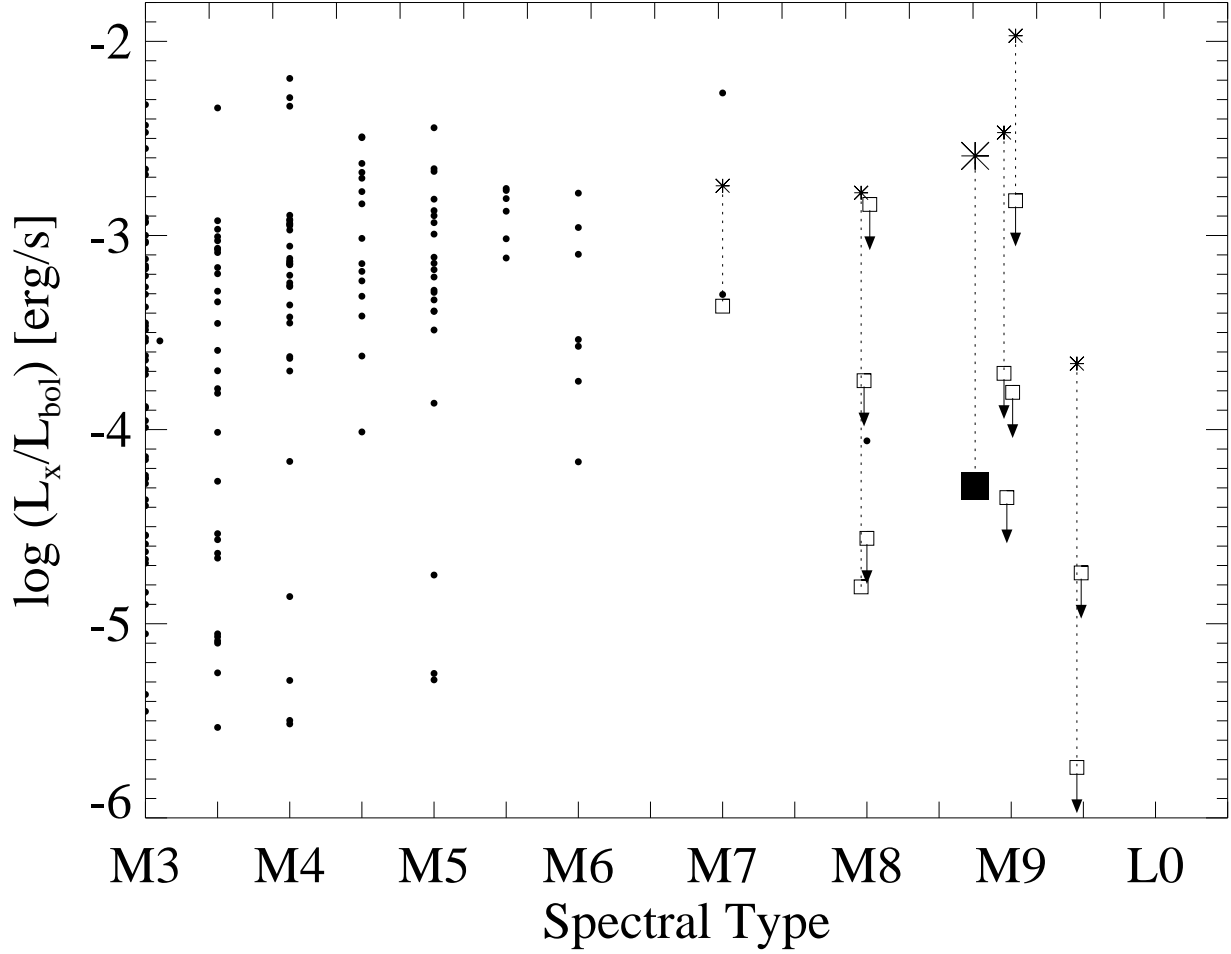}
\figcaption{Ratio of $L_{\rm x}/L_{\rm bol}$ versus spectral type for VLM field dwarfs.
{\em filled circles} -- M~dwarfs detected during the RASS (see H\"unsch et al. 1999);
{\em asterisks} -- flares, {\em squares} -- quiescent emission or upper limits (data from F93, F95, N99, F00, R00, SL02, F03, H04, S04); {\em large asterisk and square} - new data for Gl\,569\,Bab. For clarity VLM dwarfs with the same spectral type have a small offset with respect to each other on the horizontal axis.}
\label{fig:lxlbol_spt}
\end{figure}

Fig.~\ref{fig:lxlbol_spt} shows the $L_{\rm x}/L_{\rm bol}$ vs. spectral type diagram
for VLM field dwarfs. Gl\,569\,Bab adds to a handful of late-M type objects that have 
shown flares, with emission levels near or above the canonical `saturation level' of $10^{-3}$.
In the second half of the {\em Chandra} observation, after the flare decayed, Gl\,569\,Bab 
is detected as a weak X-ray source. 
If the interpretation of this emission as representative for the quiescent corona 
is correct it constitutes the first detection of persistent X-ray activity on a presumed BD. 
The sharp decline in the quiescent coronal activity for ultra-cool
dwarfs conjectured by \citet{Fleming03.1} must then be questioned. 
As seen from the upper limits in Fig.~\ref{fig:lxlbol_spt}, indeed, most previous observations
of late M~dwarfs were not sensitive enough to sample the range expected from an extrapolation
of the $L_{\rm x}/L_{\rm bol}$ values of the M0...M6 dwarfs.  
However, the non-detection of any photon from Gl\,569\,Bab in the $\sim 2.2$\,ks before the flare 
casts some doubt on this view. 
It is observed occasionally that stars do not go back into their pre-flare quiescent state
immediately after an outburst, thus we may 
see some kind of `afterglow' of the flare rather than true quiescent emission. 
On the other hand, based on the post-flare emission level a total of $2.6$ source photons 
are expected prior to the flare, consistent with the observation of zero photons at the $2\sigma$ limit. 

\begin{figure}
\epsscale{1.1}
\plotone{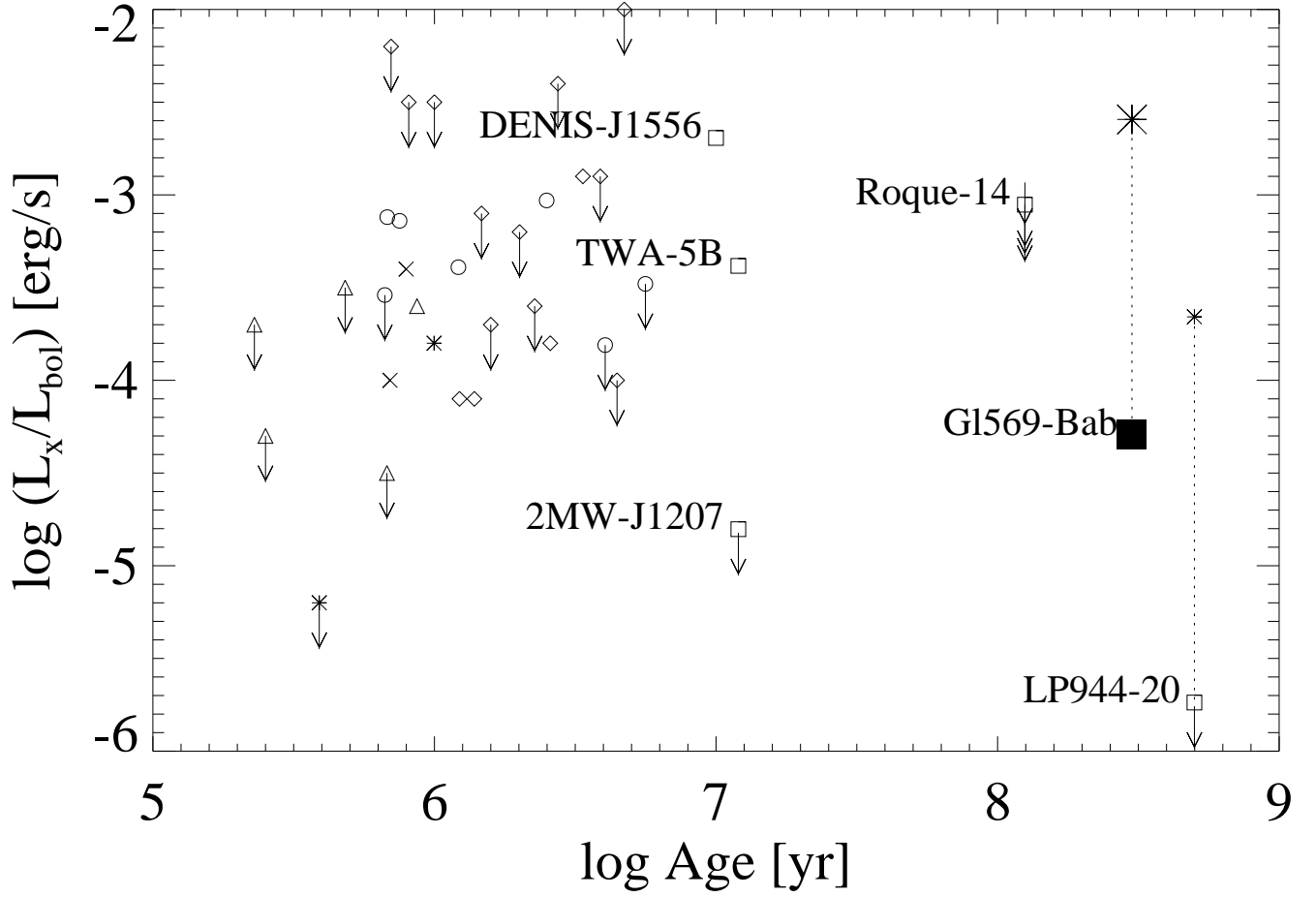}
\figcaption{Ratio of $L_{\rm x}/L_{\rm bol}$ versus age for BDs: {\em triangles, diamonds, open circles} -- bona-fide BDs in star forming regions (spectral types $\geq$\,M7) with a random one order of magnitude spread in age centered on $0.5$\,Myr for $\rho$\,Oph (X-ray data from I01), $1.5$\,Myr for IC\,348 (PZ02), and $2$\,Myr for Cha\,I (St04); flaring BDs are displayed as an asterisk, and those with accretion signatures in the H$\alpha$ profile as x-points. References for evolved BDs: DENIS\,J1556 (B04), TWA-5B (T03), 2MW-J1207 (GB04), Roque\,14 and four undetected BDs in the Pleiades (BP04), LP944-20 (R00), and Gl\,569\,Bab (this paper).}
\label{fig:lxlbol_age}
\end{figure}

The detection of a strong X-ray flare on a comparatively evolved BD demonstrates that
coronae of VLM objects can remain powerful beyond the youngest ages. 
Fig.~\ref{fig:lxlbol_age} puts the {\em Chandra} observation of Gl\,569\,Bab in context with observations
of younger BDs, and the only previous X-ray detection of a field BD, LP\,944-20. 
A factor of $4-5$ more photons have been collected for Gl\,569\,Bab with respect to all other  
late M dwarfs, allowing for the first time a meaningful spectral analysis of such an object. 
The spectrum shows significant emission above $1$\,keV, but a similarly strong cool component of
$\sim 3.2$\,MK. Low coronal temperatures seem to be characteristic for all but the youngest BDs. 
This is all the more remarkable as temperatures are known to rise in stellar flares. 
The {\em flare temperatures of the evolved BDs} (Gl\,569\,Bab and LP\,944-20) are similar to the 
{\em quiescent temperature of the middle-aged BD} TWA-5B on the one hand and to the 
{\em quiescent temperature of the evolved star} vB\,10 on the other hand. 
Upcoming observations with {\em Chandra} sampling the age and temperature space shall constrain 
the influence of these parameters on coronal emission.

\acknowledgments

I would like to thank G. Micela and E. Flaccomio for fruitful discussions and 
careful reading of the manuscript.




\clearpage

\end{document}